\documentclass[12pt,preprint]{aastex}

\begin{document}

\slugcomment{accepted to \aj\ May 13, 2007}

\title{The Mass of the Candidate Exoplanet Companion to HD 33636 from \textit{HST} Astrometry and High-Precision Radial Velocities\footnotemark}

\author{Jacob L. Bean\footnotemark, Barbara E. McArthur\footnotemark, G. Fritz Benedict\footnotemark[\value{footnote}], Thomas E. Harrison\footnotemark, Dmitry Bizyaev\footnotemark, Edmund Nelan\footnotemark, \& Verne V. Smith\footnotemark}

\footnotetext[1]{Based on data obtained with the NASA/ESA \textit{Hubble Space Telescope} (\textit{HST}) and the Hobby-Eberly Telescope (HET). The \textit{HST} observations were obtained at the Space Telescope Science Institute, which is operated by the Association of Universities for Research in Astronomy, Inc., under NASA contract NAS5-26555. The HET is a joint project of the University of Texas at Austin, the Pennsylvania State University, Stanford University, Ludwig-Maximilians-Universit\"{a}t Muenchen, and Georg-August-Universit\"{a}t G\"{o}ttingen. The HET is named in honor of its principal benefactors, William P. Hobby and Robert E. Eberly.}

\footnotetext[2]{Dept.\ of Astronomy, University of Texas, 1 University Station, C1402, Austin, TX 78712; bean@astro.as.utexas.edu}

\footnotetext[3]{McDonald Observatory, University of Texas, 1 University Station, C1402, Austin, TX 78712}

\footnotetext[4]{Department of Astronomy, New Mexico State University, 1320 Frenger Mall, Las Cruces, NM 88003}

\footnotetext[5]{National Optical Astronomy Observatory, 950 North Cherry Avenue, Tucson, AZ 85719}

\footnotetext[6]{Space Telescope Science Institute, 3700 San Martin Drive, Baltimore, MD 21218}

\footnotetext[7]{Gemini Project, National Optical Astronomy Observatory, Tucson, AZ 85719}

\begin{abstract}
We have determined a dynamical mass for the companion to HD 33636 which indicates it is a low-mass star instead of an exoplanet. Our result is based on an analysis of \textit{Hubble Space Telescope} (\textit{HST}) astrometry and ground-based radial velocity data. We have obtained high-cadence radial velocity measurements spanning 1.3 years of HD 33636 with the Hobby-Eberly Telescope at McDonald Observatory. We combined these data with previously published velocities to create a  data set that spans nine years. We used this data set to search for, and place mass limits on, the existence of additional companions in the HD 33636 system. Our high-precision astrometric observations of the system with the \textit{HST} Fine Guidance Sensor 1r span 1.2 years. We simultaneously modeled the radial velocity and astrometry data to determine the parallax, proper motion, and perturbation orbit parameters of HD 33636. Our derived parallax, $\pi_{abs}$ = 35.6 $\pm$ 0.2 mas, agrees within the uncertainties with the \textit{Hipparcos} value. We find a perturbation period $P$ = 2117.3 $\pm$ 0.8 days, semimajor axis $a_{A}$ = 14.2 $\pm$ 0.2 mas, and system inclination $i = 4\fdg1 \pm 0\fdg1$. Assuming the mass of the primary star $M_{A}$ = 1.02 $\pm$ 0.03 $M_{\sun}$, we obtain a companion mass $M_{B}$ = 142 $\pm$ 11 $M_{Jup}$ = 0.14 $\pm$ 0.01 $M_{\sun}$. The much larger true mass of the companion relative to its minimum mass estimated from the spectroscopic orbit parameters ($M \sin i$ = 9.3 $M_{Jup}$) is due to the near face-on orbit orientation. This result demonstrates the value of follow-up astrometric observations to determine the true masses of exoplanet candidates detected with the radial velocity method.
\end{abstract}


\section{INTRODUCTION}
Mass is the most important physical property of exoplanets because it is the one parameter that can establish a star's companion as an actual planet rather than a brown dwarf or low-mass star. Exoplanet masses also provide important constraints for theoretical models. As for stars, mass critically determines most of the instantaneous characteristics and long-term evolution of a planet. Therefore, we can test, and ultimately improve, our understanding of planet formation and evolution by comparing model predictions with the observed physical and orbital properties of exoplanets with measured masses.

Currently, fewer than 10$\%$ of the more than 200 candidate exoplanets\footnote{A regularly updated list of reported exoplanets can be found at http://exoplanet.eu/.} orbiting nearby stars have precisely determined masses. The rest only have known minimum masses. This is because the most successful technique for detecting candidate exoplanets, the radial velocity method, cannot be used to remove the degeneracy between the mass and orbital inclination for most of the known exoplanet candidates. In principle, radial velocities alone can be used to determine the masses of exoplanets in multi-planet systems where two or more planets are experiencing significant mutual gravitational interactions on short timescales \citep[e.g.][]{nauenberg02, rivera05}. However, only one such system is known (GJ 876), results from different groups vary significantly, and the effects of non-coplanarity have yet to be considered. Thus, establishing precise masses, rather than arguing statistically, for the majority of exoplanet candidates requires observations with complementary techniques. The techniques that have been employed to break the mass -- inclination degeneracy in radial velocity data are astrometry \citep[e.g.][]{benedict02a} and transit \citep[e.g.][]{henry00} observations. In this paper we report our determination of a dynamical mass for the exoplanet candidate companion to HD 33636 based on combined modeling of radial velocity and astrometric measurements of its perturbation orbit.

HD 33636 (= G 97-25, HIP 24205) is G0 V star at a distance of 28.7 pc \citep{perryman97}. \citet{vogt02} reported the discovery of a candidate planetary mass companion orbiting HD 33636 in a long period and eccentric orbit based on its radial velocity variations. \citet{perrier03}, and later \citet{butler06}, refined the companions's orbit parameters based on additional velocity measurements. \citet{butler06} found the spectroscopic orbit parameters period $P$ = 2128 days, eccentricity $e$ = 0.48, and velocity semiamplitude $K$ = 164 m s$^{-1}$, which resulted in an estimated minimum mass $M \sin i$ = 9.3 $M_{Jup}$ for the companion.

Assuming the \citet{butler06} orbit parameters, the corresponding minimum astrometric perturbation, $a_{A} \sin i$, of HD 33636 due to its low-mass companion would be 1.0 mas. This suggested that the perturbation from even an edge-on system orientation ($i = 90\degr$) would be detectable using the \textit{Hubble Space Telescope} (\textit{HST}) Fine Guidance Sensors (FGS),  and was the motivation for this work. \citet{benedict02a}, \citet{mcarthur04}, and \citet{benedict06} have previously used the \textit{HST} FGS in combination with radial velocity measurements obtained with ground-based telescopes to directly determine the masses of three exoplanets. Our observations and analysis to determine the mass of HD 33636's companion are similar and described in the following sections. 

In \S 2 we describe our radial velocity measurements of HD 33636 with the Hobby-Eberly Telescope (HET), which we have used to supplement the previously published velocity data for this object and search for additional low-mass, short-period companions. In \S 3 we describe our \textit{HST} observations and astrometry measurements. Our reference star spectroscopy, photometry, resulting spectrophotometric parallaxes, and estimated proper motions are discussed in \S 4. In \S 5 we present our simultaneous modeling of the radial velocity and astrometry data. From this modeling we determine the perturbation orbit parameters, which allow us to calculate the mass of the companion to HD 33636. In \S 6 we show our analysis of the radial velocity residuals from the single companion model and present a calculation of our detection limits for additional companions. We discuss the implications of our result in \S 7 and conclude in \S 8 with a summary.

\section{RADIAL VELOCITY DATA}
The ultimate precision and accuracy of our measurement of the mass for HD 33636's unseen companion depends strongly on the quality and quantity of the radial velocity data used. Unaccounted for, but detectable, additional companions or poorly constrained spectroscopic orbit parameters for the known companion would each cause a systematic error in our result. We therefore carried out high-cadence spectroscopic observations for radial velocity measurements of HD 33636 with the HET. We used these data to supplement the previously published velocities for this object and the combined data set of radial velocities spans just over nine years.

\subsection{HET Spectroscopic Observations}
We observed HD 33636 on 65 nights using the HET to feed the High Resolution Spectrograph \citep[HRS,][]{tull98} between UT dates September 20, 2005 and January 21, 2007. The HET is queue, rather than classically, scheduled, and the nights were randomly distributed throughout the observing seasons that the star was available. The HRS was used in the resolution R = 60,000 mode with a 316 gr mm$^{-1}$ echelle grating. The cross-dispersion grating was positioned so that the central wavelength of the order that fell in the break between the two CCD chips was 5936 \AA. A temperature controlled ($T = 70\fdg0 \pm 0\fdg1$ C) iodine cell (HRS3) was inserted in front of the spectrograph slit entrance for all exposures to imprint lines that provided a contemporaneous wavelength scale and instrumental profile (IP) fiducial. Three separate exposures were taken within 15 minutes on all but a few nights. In total, 195 ``target'' observations of HD 33636 were made, including 64 sets of three observations each and three additional solitary exposures. Exposure times were nominally 120 s, but varied up to twice that occasionally to account for increased seeing and/or cloud cover. 

Additionally, we observed HD 33636 once on December 6, 2006 without the iodine cell and with the same instrument setup, but in the R = 120,000 mode. The exposure time for that observation was 600 s. We used the spectrum from this ``template'' observation as the model template in the radial velocity analysis of the target spectra described in \S 2.2. 

CCD reduction and optimal order extraction were carried out for all the individual spectra using the REDUCE package \citep{piskunov02}. The final median signal-to-noise (S/N) per pixel of the target spectra is 89 and the template spectrum S/N per pixel is 188. For the CCD binning we used (two pixels in the cross-dispersion direction and one in the echelle dispersion direction), there are roughly four pixels per resolution element in the R = 60,000 mode spectra and two in the R = 120,000 spectrum.

\subsection{Radial Velocity Analysis}
We used an independent adaptation of the canonical spectrum modeling technique described by \citet{valenti95} and \citet{butler96} to measure relative radial velocities in the extracted target spectra. The observed spectra between 5020 and 5860 \AA\ were broken up into 637 separate 100 pixel ``chunks.'' Each chunk was modeled as the product of a high-resolution FTS spectrum of the iodine cell\footnote{The FTS spectrum of the HRS3 iodine cell is available at ftp://nsokp.nso.edu/FTS\_cdrom/FTS50/001023R0.004.} and a template stellar spectrum convolved with an IP. The template spectrum was Doppler shifted before being multiplied by the iodine spectrum, and this constituted the measurement of the star's relative radial velocity. Additionally, the wavelength scale of the observations was simultaneously modeled as a second order polynomial. The reduced observed spectra have an arbitrary normalization and the model normalization was a free parameter as well.

We used the sum of 11 Gaussians, one central and five satellites on each side, to represent the HRS IP. The variable parameters were the width, $\sigma$, of the central Gaussian and the heights of the satellites. The height of the central Gaussian was calculated from the $\sigma$ with the standard formula ($h = 1/2\pi\sigma$). The IP of the HRS varies considerably in shape and size along each echelle order. Therefore, we used a dynamic spacing of the satellites from the center of the central Gaussian. The spacing was set so that there were four satellites per the FHWM of the central Gaussian ($\Gamma \simeq 2.355\sigma$). The $\sigma$ of the satellites were also set dynamically. They were equal to the $\sigma$ of the central Gaussian multiplied by 0.35. Including the Doppler shift, three parameter wavelength scale, 11 parameter IP description, and spectrum normalization there was a total of 16 free parameters in the model for each spectrum chunk. 

We based our model template spectrum on the R = 120k HD 33636 spectrum taken without the iodine cell. We modeled a flat field spectrum with the iodine cell that was obtained using the same instrument setup as the template observation, taken immediately after it in the same manner as for the velocity measurements, but without the Doppler shift parameter (15 free parameters). This yielded an accurate wavelength scale and IP description as a function of the spectral position on the CCD at that time. We used this information to estimate the intrinsic spectrum of HD 33636 by removing the IP from the template observation. We used a modified Jansson technique \citep{gilliland92} for the deconvolution. This estimate of the HD 33636 intrinsic spectrum was used as the model template for the velocity measurement of the target spectra.

The spectral modeling yielded a velocity measurement for each of the 637 chunks in a spectrum. For each target spectrum, we calculated the weighted mean of the velocities from the chunks in a spectrum to give a single measured velocity and uncertainty. We weighted the velocities for each chunk according to the velocity error computed according to the formula for the intrinsic Doppler error in a section of stellar spectrum given by \citet{butler96}. The median uncertainty in our 195 measured weighted mean velocities is 2.1 m s$^{-1}$. We converted the velocity determined from each spectrum into a relative radial velocity by correcting each measurement for the barycentric motion of the observatory in the line-of-sight direction using the JPL ephemeris DE405\footnote{The JPL ephemeris data may be obtained on the Internet at ftp://ssd.jpl.nasa.gov/pub/eph/export/DE405/}. 

The RMS deviation of the HET velocities from the perturbation orbit we determine (see \S 5) is 6.1 m s$^{-1}$. Using the method of \citet{wright05}, \citet{butler06} estimated the intrinsic radial velocity ``jitter'' due to variations of the stellar photosphere for HD 33636 to be 5.2 m s$^{-1}$. Adding this value in quadrature to our velocity uncertainties we find $\chi_{\nu}^{2}$ = 1.1 for the fit to the orbit, which indicates general agreement between our radial velocity measurement and error estimation technique and that of the California -- Carnegie Planet Search group. 

Our scheme of taking three successive exposures over the course of $\sim$ 10 minutes during most nights gives us the opportunity to somewhat reduce the impact of short-term stellar noise and random errors on our orbit analysis. In addition to calculating a simple weighted mean, we also used the program \textit{Gaussfit} \citep{jefferys88} to calculate a robust weighted midpoint time, velocity, and uncertainty for each of the 64 three observation sets. Comparing the resulting data to the fit for the weighted and robust mean methods, we find RMSs of 4.4 and 3.3 m s$^{-1}$ respectively. Because of the higher quality of the robust combined velocities, we ultimately adopted those data in the analysis presented in \S 5. The velocity values, velocity uncertainties, and the corresponding heliocentric Julian dates are given in Table ~\ref{tab:table1}. The velocities we measured are relative to an arbitrary zero-point, and we have subtracted a constant value determined during our orbit analysis. Therefore, the velocities given are relative to the HD 33636 system barycenter under the assumption that our single-companion model for the system is correct.

\subsection{Total Radial Velocity Data Set}
In addition to our own measurements from the HET, we included published high precision radial velocity measurements of HD 33636 from \citet[][``Elodie'' sample]{perrier03} and \citet[][``Lick'' and ``Keck'' samples]{butler06}. G. Marcy (private communication) provided us with an updated version of the Keck velocities published in \citet{butler06}. These updated velocities result from a reanalysis of old spectra with an improved version of the California -- Carnegie Planet Search Doppler analysis software and include an additional epoch of data taken since publication that overlaps with our HET observations. Table ~\ref{tab:table2} lists the source, time coverage, number of data points, and RMS deviation from the final orbit fit for the individual samples.

The complete data set contains radial velocities from four different telescope and instrument combinations and has a time baseline of 3289 days (9.01 years). The preexisting data were a crucial component of our analysis, because HD 33636's companion has an orbital period more than four times the time span of our HET observations. However, the use of a heterogeneous data set does require particular care, because the velocities in each sample are relative to a different zero point. We found in the course of our analysis (see \S 5) that a simple offset parameter for each sample was sufficient to correct the data to the same frame of reference.

\section{\textit{HST} ASTROMETRY DATA}
We used the \textit{HST} Fine Guidance Sensor 1r (FGS1r) to carry out astrometric observations of HD 33636 and five reference stars between UT dates August 20, 2005 and October 26, 2006. A detailed overview of the FGS1r as a science instrument was given by \citet{nelan03}. Our data acquisition and reduction follow the procedure outlined by \citet{benedict00} as for the FGS3. We used the FGS1r for the current study because it provides superior fringes from which to obtain stellar positions \citep{mcarthur02}.

Table ~\ref{tab:table3} lists the log of the \textit{HST} observations. The data span 432 days (1.18 years) and include 18 epochs. Each epoch contains 2 -- 4 positional measurements of HD 33636 and the references stars, which were acquired contiguously over a time span of 26 -- 35 minutes. The observation time listed is for the midpoint of each epoch. The field was observed at multiple spacecraft roll values, and HD 33636 had to be placed in different non-central locations within the field of view (FOV) to accommodate the distribution of reference stars. The F5ND neutral density filter was used for the observations of HD 33636, while the F583W filter was used for the reference stars due to their being much fainter. To account for using a different filter for the reference stars, we included a cross-filter correction term \citep{benedict02b} in the astrometry model (see \S 5) for HD 33636.

\section{ASTROMETRIC REFERENCE STAR DATA}
Because of the high sensitivity of the FGS as an astrometer, and despite the relatively large distance of the astrometric reference stars, we had to take into account their parallaxes and proper motions in our model (see \S 5). To establish these, we determined spectrophotometric parallaxes, and adopted proper motions that could be input as constraints to our model. Additionally, our model requires input $(\bv)$ colors for all the stars to correct for chromatic aberration (lateral color). Our method for estimating reference star spectrophotometric parallaxes is discussed extensively in \citet{benedict07} and we followed the same approach in the present study. 

We obtained classification spectra of our astrometric reference stars with the R-C Spectrograph on the Cerro Tololo Inter-American Observatory (CTIO) Blanco 4 m telescope\footnote{CTIO is operated by the Association of Universities for Research in Astronomy Inc. (AURA), under a cooperative agreement with the National Science Foundation (NSF) as part of the National Optical Astronomy Observatories (NOAO).}. The spectral types were determined from these spectra by a combination of template matching and line ratios and are generally better than $\pm$ 2 subclasses. 

We also obtained $BVI$ photometry for the reference stars using the New Mexico State University 1m telescope and adopted $JHK$ magnitudes from the Two Micron All-Sky Survey (2MASS) Point Source Catalog \citep{cutri03}. We estimated the star's luminosity classes with this photometry and the reduced proper motion method. We estimated $V$ band extinctions, $A_{V}$, for the reference stars by comparing their measured colors with the expected colors for their spectral type taken from \citet{cox00}.

We next estimated $M_{V}$ values for the reference stars by assuming the prototypical values for their spectral types and luminosity classes given by \citet{cox00}. We calculated their spectrophotometric parallaxes from their measured $V$ and the estimated $A_{V}$ and $M_{V}$ values. The determined spectral types and luminosity classes, measured $V$ and $(\bv)$ values, estimated $A_{V}$ values, assumed $M_{V}$ values, and estimated spectrophotometric parallaxes for the reference stars are given in Table ~\ref{tab:table4}. Proper motions for these stars were taken from the 2nd data release of the USNO CCD Astrograph Catalog \citep[UCAC2,][]{zacharias04}. These proper motions, along with the spectrophotometric parallaxes and measured $(\bv)$ values, enter our model as observations with error.

\section{SIMULTANEOUS RADIAL VELOCITY AND ASTROMETRY SOLUTION}
We modeled the radial velocity and astrometry data simultaneously to determine the parallax, proper motion, and complete set of perturbation orbit parameters for HD 33636. With these determined parameters, we then calculated the mass of its companion. The method we used is very similar to that previously employed by \citet{benedict02a}, \citet{mcarthur04}, and \citet{benedict06} to determine the same parameters for other exoplanet host stars and their companions. 

The astrometric model for HD 33636 that we used is represented by four solved equations of condition. They are 

\begin{equation} 
x' = x + LC_{x}( \bv ) - XF_{x},
\end{equation}
\begin{equation} 
y' = y + LC_{y}( \bv ) - XF_{y},
\end{equation}
\begin{equation} 
\xi = Ax' + By' + C - P_{\alpha}\pi - \mu_{\alpha}\Delta t - ORBIT_{\alpha},
\end{equation}
\begin{equation} 
\eta = -Bx' + Ay' + F - P_{\delta}\pi - \mu_{\delta}\Delta t - ORBIT_{\delta}.
\end{equation}
Identifying terms, $x$ and $y$ are the measured coordinates; (\bv) is the photometric color; $LC_{x}$ and $LC_{y}$ are the lateral color corrections; $XF_{x}$ and $XF_{y}$ are the cross-filter corrections; A and B are plate scale and rotation parameters; $C$ and $F$ are offsets; $\mu_{\alpha}$ and $\mu_{\delta}$ are proper motion components, $\Delta t$ is the time difference from the mean epoch; $P_{\alpha}$ and $P_{\delta}$ are parallax factor components; $\pi$ is the parallax; $ORBIT_{\alpha}$ and $ORBIT_{\delta}$ are the astrometric components of the perturbation orbit; and $\xi$ and $\eta$ are the standard coordinates. The astrometric orbit is a function of the orbital parameters period ($P$), time of periastron passage ($T_{P}$), eccentricity ($e$), semimajor axis ($a_{A}$), position angle of the ascending node ($\Omega$), inclination ($i$), and the longitude of periastron passage ($\omega$). 

We also modeled the astrometry for the five reference stars in parallel with HD 33636. The equations of condition for those stars are the same as given in Equations (1 -- 4) minus the cross-filter corrections and perturbation orbit motion. The plate scale and rotation parameters and the offsets are the same for HD 33636 and the reference stars at each epoch. The parallax and proper motion are unique for each star. 

The radial velocity model for HD 33636 is given by a fifth solved equation of condition,
\begin{equation} 
\gamma = RV + G_{S} - ORBIT_{R},
\end{equation}
where $RV$ is the measured relative radial velocity; $G_{S}$ is the velocity offset for each of the four velocity samples described in \S 2.3; $ORBIT_{R}$ is the radial component of the orbital velocity; and $\gamma$ is the adopted velocity of the HD 33636 system barycenter. The exact choice of $\gamma$ is arbitrary and immaterial because our analysis is based on the relative motion of HD 33636. The important point is that its value is the same for each radial velocity sample and the $G_{S}$ values are used to correct the sample velocities to the same frame of reference. The radial velocity orbit depends on the parameters $P$, $T_{P}$, $e$, and $\omega$, which are the same for the astrometric and radial velocity orbit models. The radial velocity orbit is also dependent on the velocity semiamplitude ($K_{A}$), which does not influence the astrometric orbit. 

In addition to the shared orbit parameters, we enforced a relationship between the astrometric and radial velocity models using a constraint from \citet{pourbaix00},
\begin{equation} 
\frac{a_{A} \sin i}{\pi_{abs}} = \frac{P K_{A} \sqrt{1 - e^{2}}}{2 \pi (4.7405)},
\end{equation}
where astrometric only quantities are on the left. Quantities determined primarily or only by the radial velocities are on the right.

We solved for HD 33636's cross filter correction, coordinates, parallax, proper motion, and perturbation orbit parameters, the reference stars' coordinates, parallaxes and proper motions, the plate parameters for each astrometry epoch, lateral color corrections, and the velocity sample offsets by fitting the above models to the radial velocity and astrometry data. The HD 33636 cross-filter correction, lateral color corrections, and the reference star spectrophotometric parallaxes and proper motions were input into the model as observations with error. We used the \textit{Gaussfit} program with robust estimation and the ``fair'' metric to determine the parameter values that gave the lowest $\chi^{2}$ between our model and the measured data. We adopted the uncertainties returned by \textit{Gaussfit}, which were generated from a maximum likelihood estimation that is an approximation to a Bayesian maximum a posteriori estimator with a flat prior \citep{jefferys90}. 

\begin{figure}
\epsscale{0.8}
\plotone{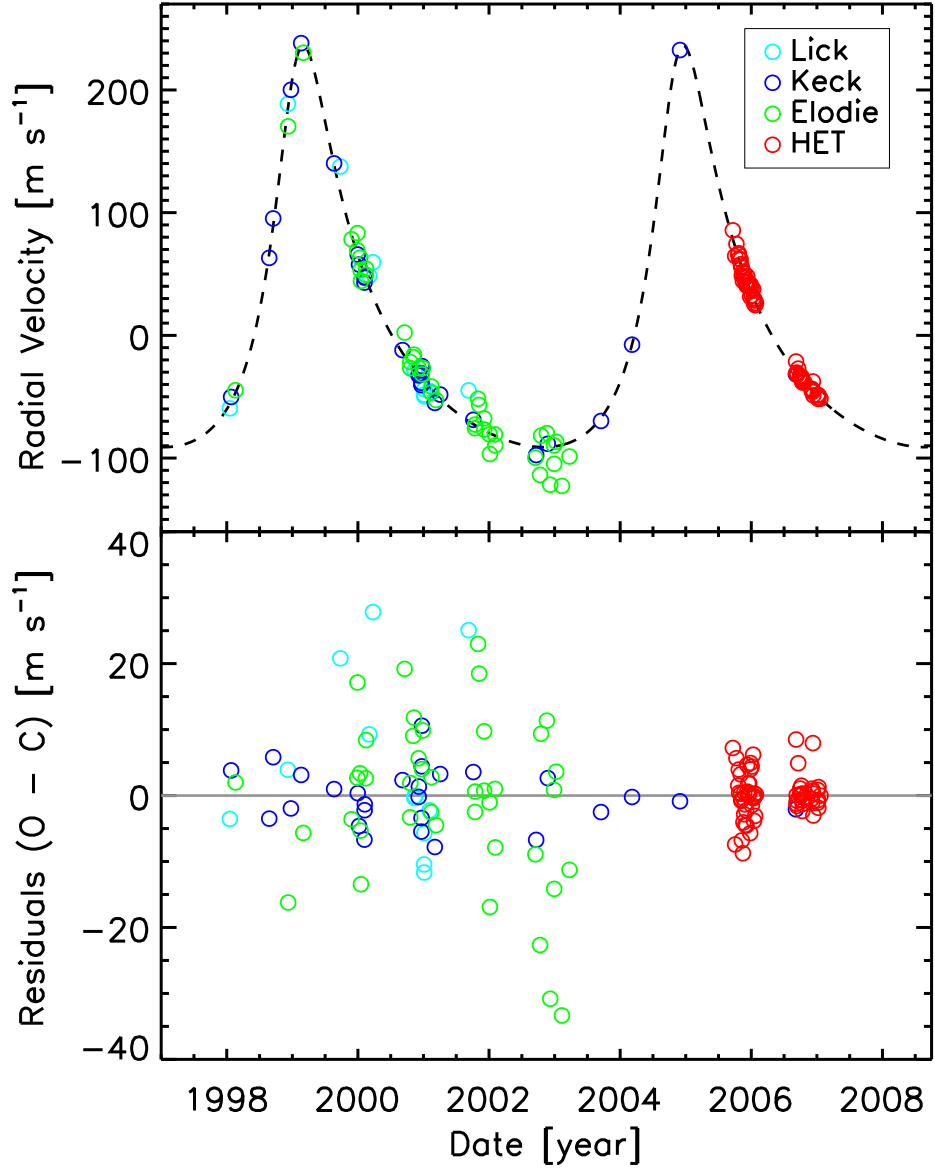}
\caption{Upper panel: Radial velocities (points) as function of time and the determined orbit (dashed line) from the simultaneous fit to the radial velocity and astrometry data. The error bars are omitted for clarity. Bottom Panel: Residuals from the fit (points).}
\label{fig:f1}
\end{figure}

The radial velocity data, our best fit, and the fit residuals as a function of time are plotted in Figure ~\ref{fig:f1}. The radial velocities for each sample in the Figure have been corrected to the system barycenter using the corresponding offset values determined in the analysis. The RMS deviations for the individual samples are given in Table ~\ref{tab:table2}. We find RMS residuals for the Keck and HET velocities of 4.2 and 3.3 m s$^{-1}$ respectively. 

\begin{figure}
\epsscale{0.8}
\plotone{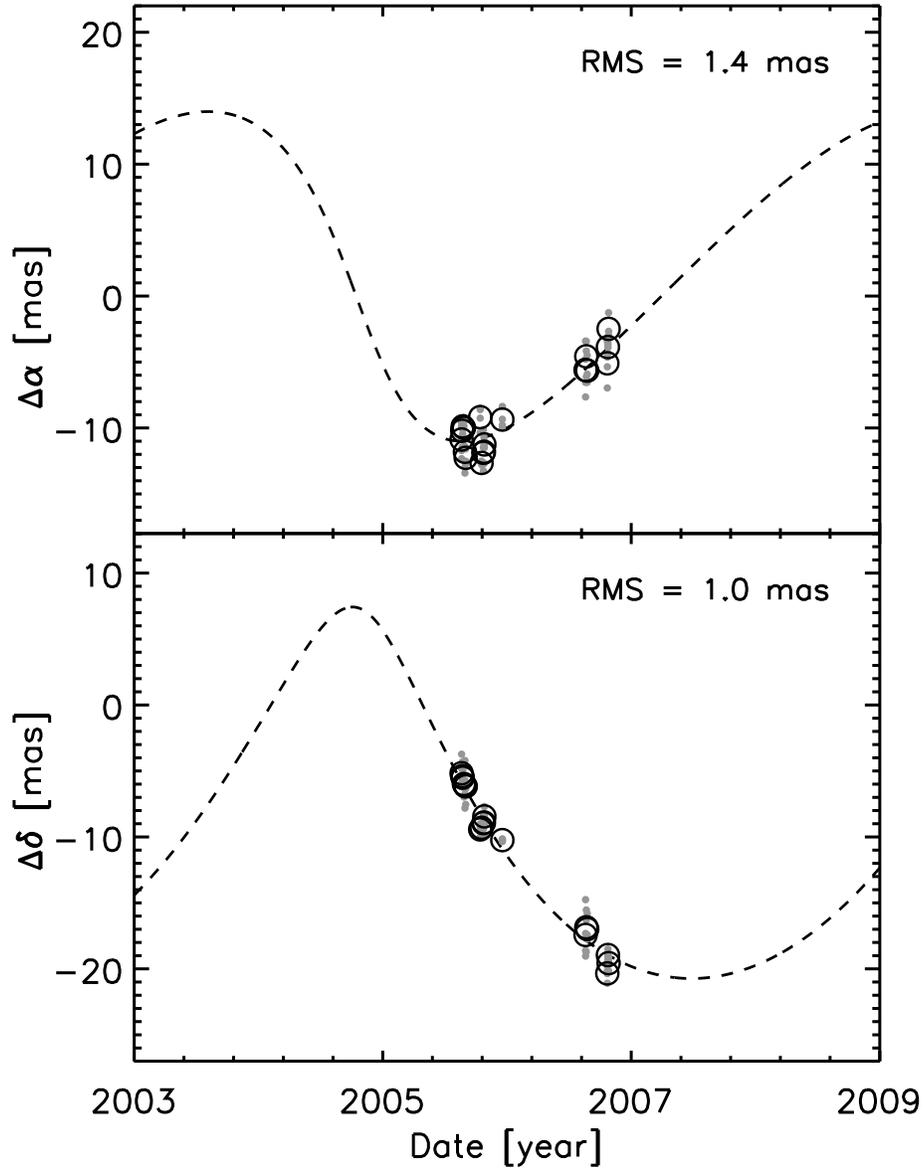}
\caption{Right ascension ($\alpha$) and declination ($\delta$) components of HD 33636's perturbation orbit as a function of time (points) and the best fit (dashed line) from the simultaneous modeling of the radial velocities and astrometry. The dots are the individual observations and the circles are per epoch (single \textit{HST} orbit) normal points consisting of 2 -- 4 individual observations.}
\label{fig:f2}
\end{figure}

\begin{figure}
\epsscale{1.0}
\plotone{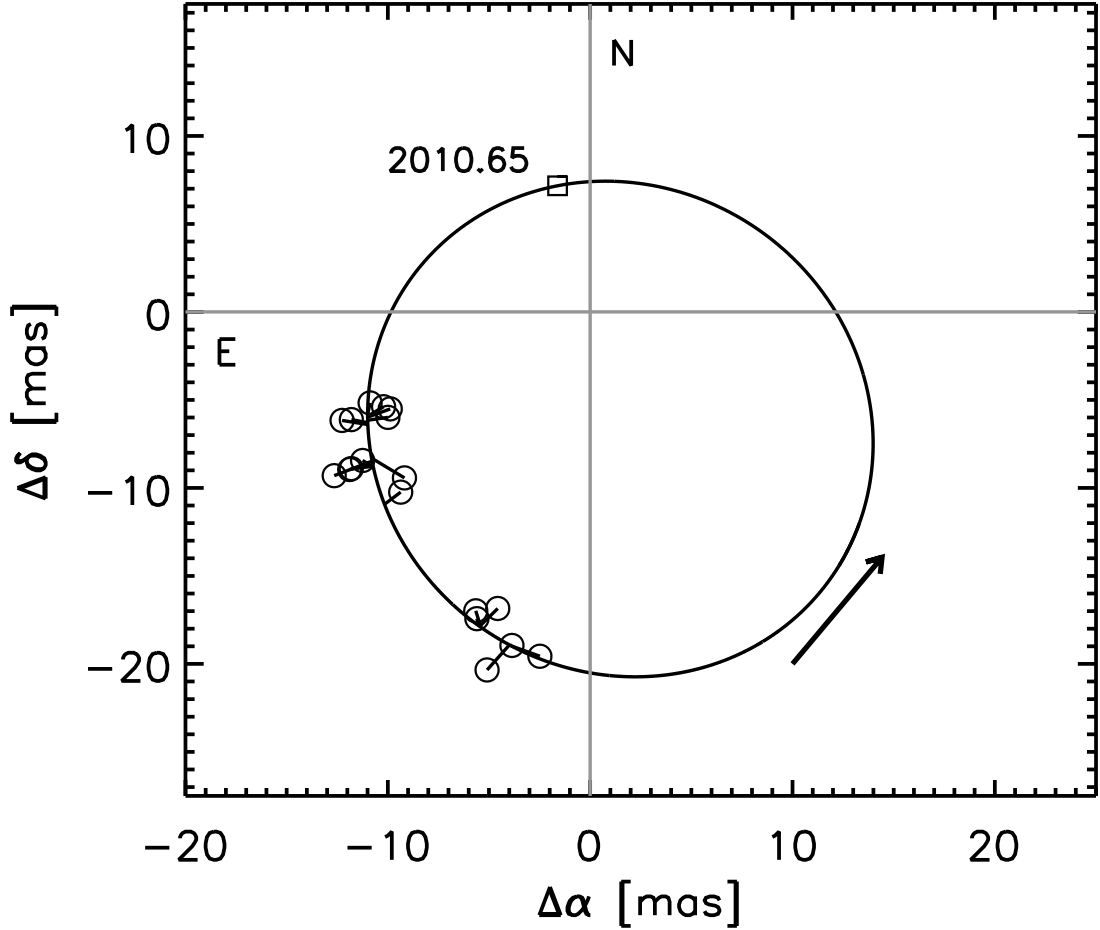}
\caption{Perturbation orbit of HD 33636 on the sky (line). The open circles are the \textit{HST} epoch normal points and are connected to the derived orbit by residual vectors. The \textit{HST} data cover 20$\%$ of the orbit period. The orbital motion direction is indicated by the arrow. The square marks the location of periastron passage and the next time of occurrence is labeled.} 
\label{fig:f3}
\end{figure}

The right ascension and declination component astrometry data as a function of time with the plate scale variation, parallactic motion, and proper motion removed are shown in Figure ~\ref{fig:f2}. These data illustrate the astrometric orbital motion of HD 33636. The fit to the data is indicated and the RMS residuals from our fit for HD 33636 are 1.4 mas and 1.0 mas in the right ascension and declination directions respectively. These values are consistent with the previously characterized FGS1r per observation precision of $\sim$ 1 mas and the median residual RMSs for the 5 reference stars, which is 1.3 mas. The quality of the simultaneous orbit fit to the astrometry and radial velocity data is evidence of the clear detection of HD 33636's perturbation due to the known companion. The same data from Figure ~\ref{fig:f2} are also shown as in the flat plane of the sky in Figure ~\ref{fig:f3}. The direction of orbital motion and the location and time of the next periastron passage (2010.65) are also indicated in Figure ~\ref{fig:f3}.

Relative coordinates and our determined absolute parallaxes and proper motions for HD 33636 and the five reference stars are given in Table ~\ref{tab:table5}. A summary of the \textit{HST} astrometry for HD 33636 is given in Table ~\ref{tab:table6}. For HD 33636 we find $\pi_{abs}$ = 35.6 $\pm$ 0.2 mas, $\mu_{\alpha}$ = 169.0 $\pm$ 0.3 mas, and $\mu_{\delta}$ = -142.3 $\pm$ 0.3 mas. Our parallax value is in excellent agreement with, but more precise than, the \textit{Hipparcos} value $\pi_{abs}$ = 34.9 $\pm$ 1.3 mas \citep{perryman97}. However, our determined proper motion is significantly different than the \textit{Hipparcos} values $\mu_{\alpha}$ = 180.8 $\pm$ 1.1 mas and $\mu_{\delta}$ = -137.3 $\pm$ 0.8 mas \citep{perryman97}. The reason for this is that the \textit{Hipparcos} model did not account for the large perturbation orbit of HD 33636. As discussed by \citet{black82}, the proper motion parameters can absorb orbital motion in astrometry data. The \textit{Hipparcos} satellite only made 16 measurements of HD 33636's position with a reported precision of better than 5 mas over a period of 2.1 years. The orbital motion of HD 33636 during this time led to the underestimation of southerly proper motion component and overestimation of the westerly proper motion component. The \textit{Hipparcos} measurements were not precise enough and did not span a sufficient amount of time to distinguish between proper motion and orbit curvature at the level necessary to trigger a multiplicity flag in the standard analysis \citep{hipp} 

In the case of our analysis, we have the benefit of foreknowledge about the existence of a companion due to the radial velocity variations. The radial velocity data that we modeled simultaneously with the astrometry data is very sensitive to most of the orbit parameters and carries the most weight in the determination of those parameters because of its quantity and quality. Additionally, we have five times the per observation astrometric precision and four times the number of observations than \textit{Hipparcos} for HD 33636. The \textit{Hipparcos} data are not useful to include in our analysis because that would necessitate the inclusion of additional transformation terms similar to the velocity offsets used to combine the radial velocity datasets. This would result in further degradation of the \textit{Hipparcos} precisions because there is no astrometric plate overlap between our \textit{HST} observations and those of \textit{Hipparcos}. Because of this, it would not have been useful to include the \textit{Hipparcos} measurements in our analysis.

\begin{figure}
\epsscale{1.0}
\plotone{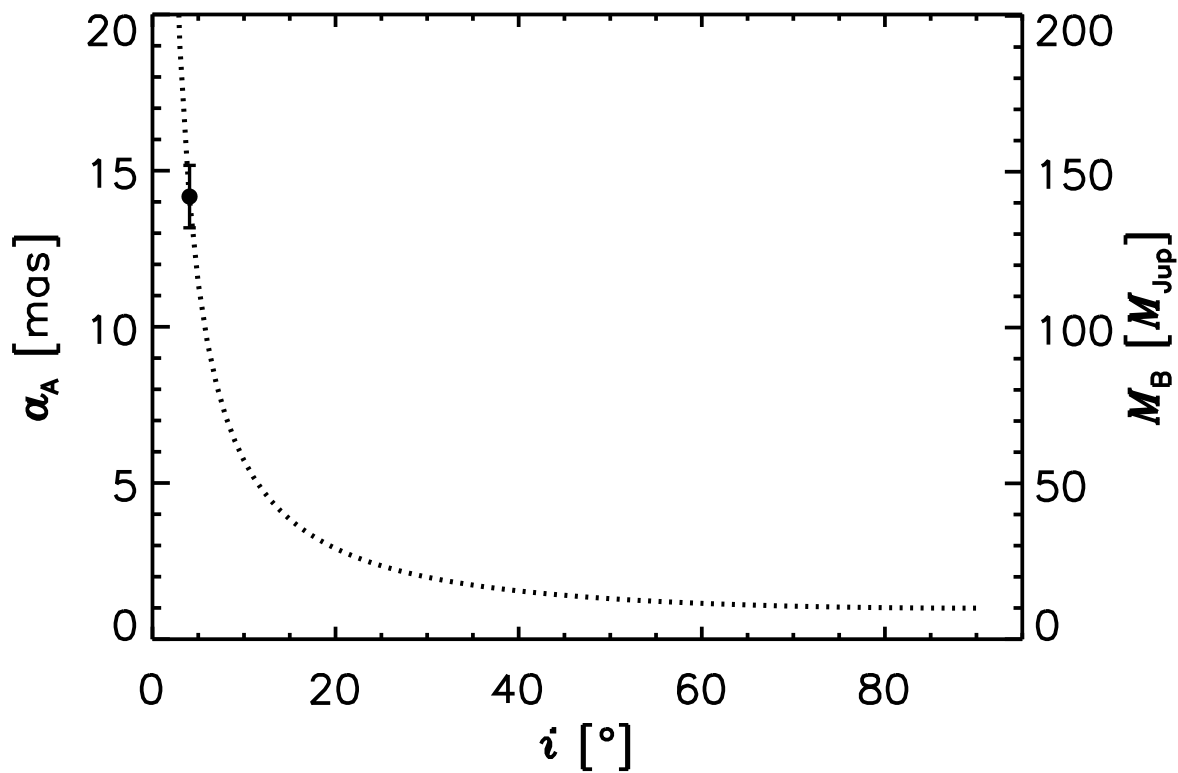}
\caption{The relationship (dotted line) between the perturbation size ($a_{A}$) and inclination angle ($i$) for fixed $P$, $K_{A}$, $e$, and $\pi_{abs}$ through the \citet{pourbaix00} relationship (eq. [6]). Our determined value for the perturbation size and inclination is given by the filled circle. The right axis maps the inclination to the corresponding companion mass ($M_{B}$). Our adopted value for the uncertainty in the companion mass is plotted as the error bar for this axis. The formal uncertainties in our determined $a_{A}$ and $i$ are smaller than the point.}
\label{fig:f4}
\end{figure}

Our derived values and uncertainties for the perturbation orbit parameters are given in Table ~\ref{tab:table7}. Figure ~\ref{fig:f4} shows a plot of the relationship between $a_{A}$ and $i$ for fixed $P$, $K_{A}$, $e$, and $\pi_{abs}$ through the \citet{pourbaix00} relationship (eq. [6]). We find that the orbit is nearly face-on, with $i = 4\fdg1 \pm 0\fdg1$ from the plane of the sky. Correspondingly, we find a large perturbation size, $a_{A}$ = 14.2 $\pm$ 0.2 mas, relative to the previously calculated minimum perturbation $a_{A} \sin i$ =  1.0 mas. We also find $P$ = 2117.3 $\pm$ 0.8 days. Assuming the mass of HD 33636 $M_{A} = 1.02 \pm 0.03 M_{\sun}$ \citep{takeda07}, we find the mass of the companion, $M_{B}$, by iterating the equation
\begin{equation} 
\bigg(\frac{a_{A}}{\pi_{abs}}\bigg)^{3} = M_{B}^{3}\bigg(\frac{P}{M_{A}+M_{B}}\bigg)^{2}.
\end{equation}
This yields $M_{B} = 142^{+3.3}_{-1.8}\ M_{Jup} = 0.136^{+0.003}_{-0.002}\ M_{\sun}$, which is also relatively large compared to the minimum mass $M \sin i = 9.3 M_{Jup}$, calculated from the spectroscopic orbit parameters. We have elected to round up to the next significant figure in solar units to account for possible systematic errors in the adopted mass of HD 33636 and other aspects of the analysis. Our final adopted companion mass uncertainty, which is also given in Table ~\ref{tab:table7}, is $0.01 M_{\sun} = 11 M_{Jup}$. We conclude from this result that HD 33636's companion is a low-mass star and not an exoplanet.

\section{LIMITS ON ADDITIONAL COMPANIONS}
\begin{figure}
\epsscale{1.0}
\plotone{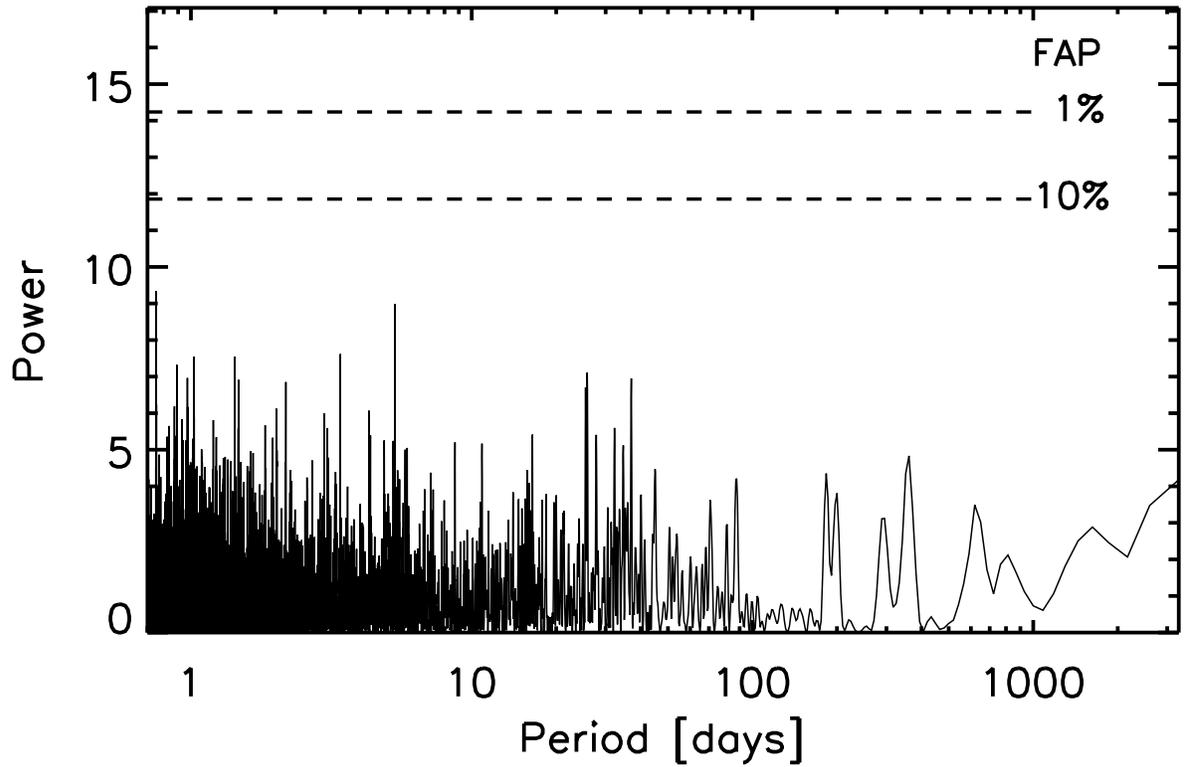}
\caption{Periodogram of the radial velocity residuals from the best fit, single-companion orbit with 1$\%$ and 10$\%$ false alarm probability (FAP) limits indicated. No periodicity is detected with FAP $<$ 70$\%$, which indicates that there are no additional companions in the system for our detection limits (see \S 6).}
\label{fig:f5}
\end{figure}

As discussed in \S 2, one motivation for obtaining additional radial velocity data beyond the previously published data was to search for, and place limits on, additional companions in the system. To do this we analyzed the radial velocity residuals from the orbit model described in \S 5. We calculated a (Lomb) periodogram \citep{press92} for the residuals from the model fit and the result is shown in Figure ~\ref{fig:f5}. We searched for prominent peaks at periods shorter than the time span of the data set (3289 days) and found none with significant power or having a $< 70\%$ false alarm probability (FAP). We also carried out this analysis on the residuals for the individual samples separately and, again, found no signs for regular periodicity. In addition, we analyzed the radial velocity data by fitting trends to the total data set and individual samples' residuals and found nothing significant. For the total data set we found a trend of 0.3 m s$^{-1}$ over the 9 year span, which is within the velocity amplitude uncertainty (0.6 m s$^{-1}$, see Table ~\ref{tab:table7}).

To quantify our detection limits for additional companions we used a method similar to that of \citet{wittenmyer06}. We adopted the velocity residuals as a noise sample for simulated orbits. We generated simulated radial velocity orbits for each of 200 period values from 0.8 to 3289 days. The discrete period values were spaced evenly on a logarithmic scale. For each period, we began by assuming a starting velocity semimajor amplitude of 4 m s$^{-1}$. We calculated 100 orbits for the given period and velocity semiamplitude and selected different time and longitude of periastron passage values with a pseudo-random number generator. We calculated the radial velocity values of the orbits at each of the dates in our measured velocity data set. We scrambled our velocity residuals, also with a pseudo-random number generator, and added them to the simulated data as noise. For each of the 100 trials we calculated a periodogram of the simulated velocity data. If the power at the period used to generate the orbit was found to correspond to a FAP $< 0.1\%$ then that was counted as a successful detection. If 99 of the 100 trials at a velocity semimajor amplitude resulted in detections then that velocity was considered the detection limit at the period value and the next period value was considered. If two or more non-detections occurred then the velocity semimajor amplitude was increased 0.5 m s$^{-1}$ and the process continued until the velocity limit for the simulated period was found. We carried out this analysis for three different values of orbital eccentricity $e$ = 0.0, 0.4, and 0.7. The upper eccentricity limit of 0.7 was chosen because $> 95\%$ of radial velocity detected exoplanet candidates have eccentricities lower than that value. For the simulations with an eccentricity of 0.0, the longitude of periastron is a meaningless parameter, and the 100 randomized values of the time of periastron served to set the relative phase.

\begin{figure}
\epsscale{0.8}
\plotone{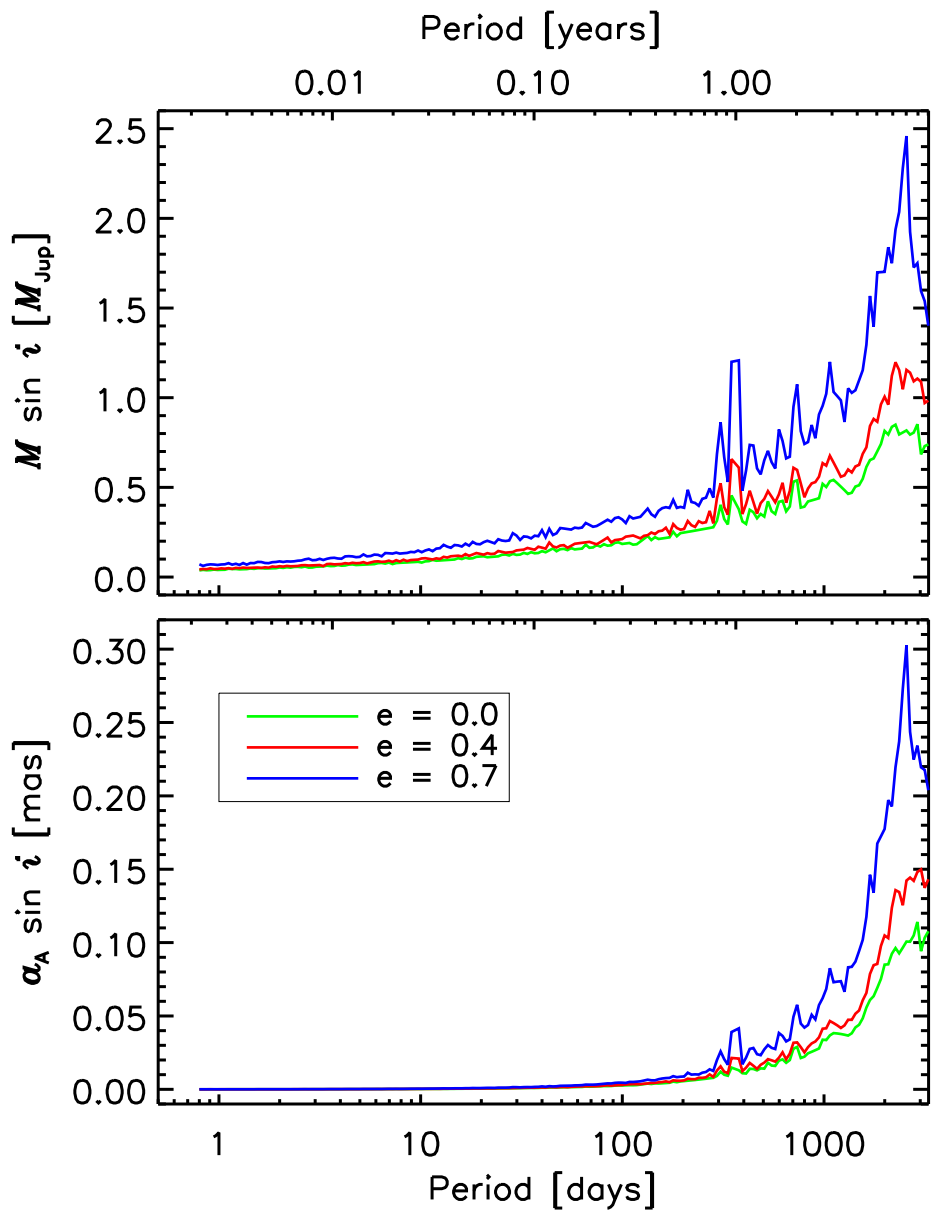}
\caption{Results from the detection limit simulations converted to hypothetical companion minimum masses ($M \sin i$, upper panel) and minimum astrometric perturbation size of HD 33636 ($a_{A} \sin i$, lower panel). The different lines represent the different assumed eccentricity values. Values above the lines would have been detected with a periodogram analysis of the radial velocity data.}
\label{fig:f6}
\end{figure}

With the set of eccentricities, periods, and velocity detection limits derived from this procedure we calculated the corresponding minimum companion masses and astrometric perturbation sizes. The results are shown in Figure ~\ref{fig:f6}. From these data we infer that our radial velocity data were at least sensitive enough to detect companions with minimum masses $M \sin i > 0.2\ M_{Jup}$, orbital periods $P <$ 100 days, and orbital eccentricities $< 0.4$. For the same orbital eccentricities, companions with $M \sin i > 1.2\ M_{Jup}$ and orbital periods up to 3300 days are ruled out. We also find that radial velocity undetected companions would have a minimum astrometric signature $a \sin i <$ 0.05 mas for $P <$ 1000 days and reasonable eccentricities, which is beyond the detection threshold of the \textit{HST} observations. Taken together, the periodogram analysis, slope analysis, and simulation of the residuals indicate no additional companions in the system, verifying that our single companion model to fit the radial velocity and astrometry data was appropriate.

\section{DISCUSSION}
HD 33636's companion is likely a M6 V star, with magnitudes $\Delta V \approx 8.0$ and $\Delta K \approx 4.0$ from the G0 V primary \citep{cox00}. From this we estimate that HD 33636 should have only a 0.03 mag enhancement in $K$ from that expected for a solitary G0 V star. The two stars' separation at apastron will be 0\farcs15. Therefore, high-resolution imaging could be useful for additional study of this system.

The mass we find for HD 33636's companion indicates that it is not a planet at the $\sim12\sigma$ level assuming the standard 13 $M_{Jup}$ upper limit for planets. This is the first definitive example of a planet candidate detected with the radial velocity method being confirmed to exist, but to have a non-planetary mass. Previously, \citet{reffert06} analyzed the \textit{Hipparcos} Intermediate Astrometric Data for HD 38529 and 168443 with radial velocity data as a constraint and found companion masses $M = 37^{+36}_{-19}\ M_{Jup}$ and $M = 34 \pm 12\ M_{Jup}$ respectively. However, these are both $< 1.8\ \sigma$ results and more observations are needed to refine these estimates.

In contrast, \citet{benedict02a}, \citet{mcarthur04}, and \citet{benedict06} have directly confirmed the planetary nature of exoplanets around GJ 876, $\rho^{1}$ Cancri, and $\epsilon$ Eridani with the same method used in this paper. The GJ 876 and $\rho^{1}$ Cancri systems contain additional, non-astrometrically detected companions. If these systems are coplanar then the additional companions are also planets. Three other planet candidates originally detected with the radial velocity method have been observed to transit their host star, HD 209458b \citep{henry00, charbonneau00}, HD 189733b \citep{bouchy05}, and HD 149026b \citep{sato05}, and thus have measured masses and are confirmed planets.

From mathematical arguments, the median value for the inclination of binary (star + star or star + planet) orbits is 60$\degr$. This is supported by the distribution of the inclinations of visual binary star orbits in the Washington Double Star Catalog \citep{mason06}, which has a broad peak in frequency centered around 60$\degr$. An inclination $i = 60\degr$ corresponds to a multiplicative factor of 1.15 to the minimum mass calculated from spectroscopic orbit parameters. Therefore, we expect that most of the candidate exoplanets detected with the radial velocity method and having minimum masses $M \sin i \la 11 M_{Jup}$ are actually planets. 

Nearly face-on orbits, like the one we have determined for the HD 33636 system, should be rare. Orbits with $i \leq 5\degr$ are expected to make up only 0.4$\%$ of an unbiased distribution. Nevertheless, our result is a striking example proving that minimum masses are not true masses and that not all of the planet candidates are actual planets. This demonstrates the value of follow-up astrometric observations and photometric monitoring for potential transits to determine the true masses of exoplanet candidates detected with the radial velocity method.

\citet{vf05} found HD 33636's iron abundance [Fe/H] = -0.13, which is consistent with the solar neighborhood average \citep{carlos04}, but on the lower end of the distribution of stars with detected planet candidate companions \citep{fv05}. Although HD 33636's companion falls outside the period range ($P <$ 4 years) considered by \citet{fv05}, it is reasonable to assume that the metallicities of host stars to high-mass planets at all periods should be distributed at similar high values as is suggested by the core accretion model of planet formation. For stars with -0.50 $\leq$ [Fe/H] $\leq$ 0.0, \citet{fv05} found a planet candidate detection rate $< 3\%$. Above solar metallicity, they found the detection rate increased, and was up to 25$\%$ for stars with [Fe/H] $>$ +0.3 dex. In this context, the fact that HD 33636 is not a planet hosting star is not as surprising as it would have been if it had [Fe/H] $\gg$ 0.0, because removing it from the sample strengthens the correlation between planets and host star metallicity.

Along the lines of planet host star abundances, \citet{ecuvillon06} and \citet{chen06} have included HD 33636 as a planet hosting star for studies of oxygen and lithium abundances respectively. \citet{ecuvillon06} found that planet hosting stars could have oxygen abundances, [O/H], enhanced from a volume limited control sample by 0.1 -- 0.2 dex, but that there was a large uncertainty in the measurements and ambiguity with the effects of galactic chemical evolution. \citet{chen06} found that planet hosting stars could have depleted lithium abundances relative to stars without detected planet companions. HD 33636 was found to be slightly oxygen rich ($\sim$ 0.1 dex) relative to control samples stars with similar [Fe/H] in the \citet{ecuvillon06} study. Conversely, \citet{chen06} found that HD 33636 showed no signs of lithium depletion as was the case in many other ostensibly planet hosting stars. 

Another study with conclusions that could be affected by our result was done by \citet{beichman05}, who looked for infrared excess due to debris disks around planet hosting stars with \textit{Spitzer}. HD 33636 was one of six planet hosting stars that showed excess emission at 70 $\mu$m and it also was one of the three that showed the most significant excess. In addition, \citet{beichman05} found six stars that did not have detected planets showing the same type of excess. Moving HD 33636 from the planet hosting sample to the non-planet hosting sample has the effect of increasing the offset in the frequency distribution for 70 $\mu$m excess between the two samples. Surprisingly, the non-planet hosting stars then have a higher infrared excess detection rate (7/60) than the planet hosting stars (6/80). This indicates there is no special correlation between planets and debris disks within the current planet and disk detection limits. Both this study, and the two mentioned above, depend on data that is difficult to measure. Resolving a true planet population from the sample of planet candidates with mass measurements will ultimately increase the impact of these and other similar studies and also permit new studies to be undertaken.

\section{CONCLUSIONS}
We have determined HD 33636's parallax, proper motion, and perturbation orbit parameters from simultaneously modeling radial velocity and astrometry data. The total radial velocity data set spans 9.0 years and includes 1.3 years of data obtained by us with the Hobby-Eberly Telescope. The astrometry data spans 1.2 years and was obtained with the FGS1r instrument on the \textit{Hubble Space Telescope}. 

Our derived parallax agrees with the \textit{Hipparcos} value within the measurement uncertainties. We find a system inclination $i = 4\fdg1 \pm 0\fdg1$, which indicates a near face-on orbit and a companion mass much larger than the minimum mass calculated from the spectroscopic orbit parameters. With our determined perturbation period $P$ = 2117.3 $\pm$ 0.8 days and semimajor axis $a_{A}$ = 14.2 $\pm$ 0.2 mas and assumed mass for HD 33636 $M_{A} = 1.02 \pm 0.03 M_{\sun}$, we obtain the mass of the companion $M_{B} = 0.14 \pm 0.01 M_{\sun}$. We conclude that HD 33636's companion is a M6 V star rather than an exoplanet. 

\acknowledgments
We are grateful to G. Marcy for providing the unpublished Keck radial velocities. In addition, we acknowledge R. P. Butler, D. A. Fischer, J. Johnson, K. Peek, S. S. Vogt, and J. T. Wright for their contributions to measuring those velocities. We also thank the anonymous referee for a careful reading of the manuscript and constructive comments which improved this paper. JLB, BEM, and GFB acknowledge support from NASA GO-09407, GO-09969, GO-09971, GO-10103, GO-10610, and GO-10989 from the Space Telescope Science Institute, which is operated by the Association of Universities for Research in Astronomy, Inc., under NASA contract NAS5-26555; and from JPL 1227563 ({\it SIM} MASSIF Key Project, Todd Henry, P.I.), administered by the Jet Propulsion Laboratory.

\clearpage
\begin{deluxetable}{cc}
\tabletypesize{\scriptsize}
\tablecolumns{2}
\tablewidth{0pc}
\tablecaption{HET Radial Velocities for HD 33636}
\tablehead{
 \colhead{HJD - 2450000.0} &
 \colhead{RV (m s$^{-1}$)} 
}
\startdata
 3633.9377 &  85.6 $\pm$ 3.2 \\
 3646.9084 &  64.7 $\pm$ 3.3 \\     
 3653.9013 &  74.5 $\pm$ 3.5 \\     
 3663.8716 &  65.8 $\pm$ 2.9 \\     
 3666.8409 &  66.9 $\pm$ 3.2 \\     
 3668.8329 &  62.5 $\pm$ 3.4 \\     
 3676.8395 &  61.8 $\pm$ 3.5 \\     
 3678.8104 &  58.0 $\pm$ 2.9 \\     
 3680.8050 &  56.1 $\pm$ 3.0 \\     
 3682.7969 &  55.4 $\pm$ 3.2 \\     
 3683.8119 &  48.8 $\pm$ 3.3 \\     
 3689.9219 &  44.3 $\pm$ 3.2 \\     
 3691.7897 &  51.9 $\pm$ 3.0 \\     
 3692.7895 &  47.8 $\pm$ 2.9 \\     
 3696.7711 &  47.4 $\pm$ 3.0 \\     
 3697.7683 &  50.3 $\pm$ 2.9 \\     
 3700.7600 &  47.3 $\pm$ 2.6 \\     
 3703.7523 &  43.0 $\pm$ 4.0 \\     
 3708.8624 &  43.2 $\pm$ 3.3 \\     
 3709.8785 &  40.5 $\pm$ 3.4 \\     
 3713.7238 &  48.2 $\pm$ 3.9 \\     
 3714.8699 &  43.7 $\pm$ 3.9 \\     
 3719.6983 &  41.3 $\pm$ 4.1 \\     
 3719.8439 &  43.1 $\pm$ 3.9 \\     
 3724.8191 &  41.2 $\pm$ 3.4 \\     
 3724.8225 &  39.7 $\pm$ 3.6 \\     
 3730.6675 &  31.5 $\pm$ 3.7 \\     
 3731.6737 &  40.8 $\pm$ 3.5 \\     
 3732.6648 &  41.4 $\pm$ 3.5 \\     
 3738.6609 &  38.7 $\pm$ 3.1 \\     
 3739.6456 &  32.5 $\pm$ 3.4 \\     
 3746.6229 &  37.7 $\pm$ 4.7 \\     
 3748.6334 &  30.4 $\pm$ 3.7 \\     
 3751.7516 &  25.9 $\pm$ 3.4 \\     
 3753.7481 &  28.9 $\pm$ 3.7 \\     
 3754.6125 &  28.6 $\pm$ 3.7 \\     
 3755.6016 &  28.7 $\pm$ 3.6 \\     
 3757.7514 &  24.6 $\pm$ 3.9 \\     
 3762.5922 &  26.2 $\pm$ 4.4 \\     
 3985.9800 & -21.3 $\pm$ 3.4 \\     
 3987.9655 & -31.4 $\pm$ 2.7 \\     
 3988.9695 & -30.3 $\pm$ 2.7 \\     
 3989.9691 & -31.3 $\pm$ 2.7 \\     
 3990.9631 & -32.3 $\pm$ 2.7 \\     
 3997.9516 & -27.1 $\pm$ 2.7 \\     
 4007.9220 & -33.6 $\pm$ 2.8 \\     
 4008.9051 & -32.9 $\pm$ 3.1 \\     
 4014.9009 & -35.0 $\pm$ 2.7 \\     
 4015.9059 & -33.7 $\pm$ 3.3 \\     
 4018.8874 & -35.4 $\pm$ 3.0 \\     
 4019.8780 & -36.6 $\pm$ 2.8 \\     
 4020.8750 & -38.4 $\pm$ 3.0 \\     
 4021.8729 & -37.5 $\pm$ 3.0 \\     
 4031.8466 & -38.2 $\pm$ 3.1 \\     
 4072.7382 & -45.0 $\pm$ 3.4 \\     
 4073.7364 & -43.6 $\pm$ 3.0 \\     
 4075.8628 & -46.1 $\pm$ 3.0 \\     
 4076.7277 & -44.4 $\pm$ 3.1 \\     
 4079.7194 & -37.6 $\pm$ 3.3 \\     
 4080.8438 & -45.4 $\pm$ 3.0 \\     
 4081.8592 & -48.9 $\pm$ 3.2 \\     
 4105.6559 & -48.5 $\pm$ 4.0 \\     
 4106.7734 & -50.6 $\pm$ 3.7 \\     
 4108.7813 & -50.5 $\pm$ 3.9 \\     
 4109.7746 & -51.8 $\pm$ 3.7 \\     
 4110.7867 & -48.8 $\pm$ 4.3 \\     
 4121.6098 & -51.6 $\pm$ 4.1 \\     
\enddata
\label{tab:table1}
\end{deluxetable}

\clearpage
\begin{deluxetable}{lccc}
\tabletypesize{\scriptsize}
\tablecolumns{4}
\tablewidth{0pc}
\tablecaption{The Radial Velocity Samples}
\tablehead{
 \colhead{Sample} &
 \colhead{Time Span} &
 \colhead{N} &
 \colhead{RMS (m\ s$^{-1}$)} 
}
\startdata
 Lick    & 1998.05 -- 2001.69 & 12 & 13.6 \\
 Keck    & 1998.07 -- 2006.68 & 27 &  4.2 \\
 Elodie  & 1998.13 -- 2003.23 & 42 & 12.2 \\
 HET     & 2005.72 -- 2007.05 & 67 &  3.3 \\
\enddata
\label{tab:table2}
\end{deluxetable}

\clearpage
\begin{deluxetable}{ccc}
\tabletypesize{\scriptsize}
\tablecolumns{3}
\tablewidth{0pc}
\tablecaption{Log of FGS1r Observations}
\tablehead{
 \colhead{HJD - 2450000.0} &
 \colhead{N\tablenotemark{a}} &
 \colhead{\textit{HST} Roll ($\degr$)} 
}
\startdata
 3603.2802 & 4 & 281.4 \\
 3605.2786 & 4 & 281.0 \\
 3606.5443 & 4 & 281.0 \\
 3610.2767 & 4 & 281.0 \\
 3613.1421 & 3 & 281.5 \\
 3615.4771 & 4 & 281.0 \\
 3658.0422 & 2 & 246.9 \\
 3662.0402 & 4 & 247.6 \\
 3667.3715 & 4 & 247.7 \\
 3669.1033 & 4 & 250.9 \\
 3670.2403 & 4 & 245.6 \\
 3723.0689 & 4 & 182.0 \\
 3967.5470 & 4 & 281.7 \\
 3969.6785 & 3 & 281.0 \\
 3972.3435 & 3 & 281.0 \\
 4031.4272 & 4 & 251.9 \\
 4033.1582 & 4 & 251.9 \\
 4035.2237 & 4 & 251.9 \\
\enddata
\tablenotetext{a}{Number of observations of HD 33636 per epoch.}
\label{tab:table3}
\end{deluxetable}

\clearpage
\begin{deluxetable}{ccccccc}
\tabletypesize{\scriptsize}
\tablecolumns{7}
\tablewidth{0pc}
\tablecaption{Astrometric Reference Star Data}
\tablehead{
 \colhead{Identification} &
 \colhead{Spectral Type} &
 \colhead{$V$} &
 \colhead{$\bv$} &
 \colhead{$A_{V}$} &
 \colhead{$M_{V}$\tablenotemark{a}} &
 \colhead{$\pi_{abs}$ (mas)} 
}
\startdata
 Ref-1 & F6 V   & 15.2 & 0.6 & 0.5 & 3.6 & 0.6 $\pm$ 0.1 \\
 Ref-2 & F6 V   & 14.1 & 0.6 & 0.4 & 3.6 & 0.9 $\pm$ 0.2 \\
 Ref-3 & K6 V   & 15.3 & 1.3 & 0.1 & 7.3 & 2.6 $\pm$ 0.5 \\
 Ref-4 & G2 V   & 13.1 & 0.6 & 0.1 & 4.6 & 2.0 $\pm$ 0.4 \\
 Ref-5 & K3 III &  9.9 & 1.3 & 0.1 & 0.3 & 1.2 $\pm$ 0.2 \\
\enddata
\tablenotetext{a}{Taken from \citet{cox00} for the measured spectral types.}
\label{tab:table4}
\end{deluxetable}

\clearpage
\begin{deluxetable}{lccccc}
\tabletypesize{\scriptsize}
\tablecolumns{6}
\tablewidth{0pc}
\tablecaption{Astrometry Catalog}
\tablehead{
 \colhead{Star} &
 \colhead{$\alpha$\tablenotemark{a}} &
 \colhead{$\delta$\tablenotemark{a}} &
 \colhead{$\pi_{abs}$} &
 \colhead{$\mu_{\alpha}$} &
 \colhead{$\mu_{\delta}$} \\
 \colhead{} &
 \colhead{(arcsec)} &
 \colhead{(arcsec)} &
 \colhead{(mas)} &
 \colhead{(mas)} &
 \colhead{(mas)} 
}
\startdata
 HD 33636 &  -97.8347 $\pm$ 0.0003 & 803.0143 $\pm$ 0.0005 & 35.6 $\pm$ 0.2 & 169.0 $\pm$ 0.3 & -142.3 $\pm$ 0.3 \\
 Ref-1    & -172.1200 $\pm$ 0.0004 & 800.5260 $\pm$ 0.0005 &  5.7 $\pm$ 0.1 &   8.6 $\pm$ 0.6 &   -9.2 $\pm$ 0.5 \\
 Ref-2    & -159.6280 $\pm$ 0.0002 & 775.5080 $\pm$ 0.0005 &  9.2 $\pm$ 0.1 &   0.6 $\pm$ 0.3 &  -12.2 $\pm$ 0.3 \\
 Ref-3    & -113.6041 $\pm$ 0.0004 & 588.7119 $\pm$ 0.0005 &  2.6 $\pm$ 0.2 &  -4.1 $\pm$ 0.6 &    4.8 $\pm$ 0.5 \\
 Ref-4    & -216.5286 $\pm$ 0.0003 & 618.6621 $\pm$ 0.0005 &  2.0 $\pm$ 0.2 &  -1.2 $\pm$ 0.4 &   -2.5 $\pm$ 0.4 \\
 Ref-5    & -262.7320 $\pm$ 0.0002 & 583.6012 $\pm$ 0.0005 &  1.2 $\pm$ 0.1 &   0.0 $\pm$ 0.2 &   -6.0 $\pm$ 0.2 \\
\enddata
\tablenotetext{a}{The right ascension and declination coordinates are relative to $\alpha = 4^{h}10^{m}22\fs66, \delta = 5\degr 13' 29\farcs0$, J2000.0}
\label{tab:table5}
\end{deluxetable}

\clearpage
\begin{deluxetable}{ll}
\tabletypesize{\scriptsize}
\tablecolumns{2}
\tablewidth{0pc}
\tablecaption{Summary of \textit{HST} Astrometry}
\tablehead{
 \colhead{Parameter} &
 \colhead{Value} 
}
\startdata
\textit{HST} study duration & 1.2 yr \\
Number of HD 33636 obsrvations & 67 \\
Number of epochs & 18 \\
Number of reference stars & 5 \\
HD 33636 $(V)$ &  7.1 \\
Reference stars $\langle (V) \rangle$ & 13.5 \\
HD 33636 $(\bv)$ & 0.6 \\
Reference stars $\langle (\bv) \rangle$ & 0.9 \\
\textit{HST} parallax\tablenotemark{a} & 35.6 $\pm$ 0.2 mas \\
\textit{Hipparcos} parallax & 34.9 $\pm$ 1.3 mas \\
\textit{HST} proper motion\tablenotemark{a} & 220.9 $\pm$ 0.4 mas yr$^{-1}$ \\
In position angle\tablenotemark{a} & 130\fdg1 $\pm$ 0\fdg1 \\
\textit{Hipparcos} proper motion & 227.0 $\pm$ 1.4 mas yr$^{-1}$ \\
In position angle & 127\fdg2 $\pm$ 0\fdg3 \\
\enddata
\tablenotetext{a}{From the simultaneous modeling of the radial velocity and astrometry data.}
\label{tab:table6}
\end{deluxetable}

\clearpage
\begin{deluxetable}{ll}
\tabletypesize{\scriptsize}
\tablecolumns{2}
\tablewidth{0pc}
\tablecaption{HD 33636 Perturbation Orbit Parameters and Companion Mass}
\tablehead{
 \colhead{Parameter} &
 \colhead{Value} 
}
\startdata
$K_{A}$ & 163.5 $\pm$ 0.6 m s$^{-1}$           \\
$P$ & 2117.3 $\pm$ 0.8 days                      \\
$T_{0}$(JD) & 2451198.3 $\pm$ 2.0              \\
$e$ & 0.48 $\pm$ 0.02                            \\
$\omega$ & 337\fdg0 $\pm$ 1\fdg6           \\
$a_{A}$ & 14.2 $\pm$ 0.2 mas                   \\
$\Omega$ & 125\fdg6 $\pm$ 1\fdg6           \\
$i$ & 4\fdg0 $\pm$ 0\fdg1                   \\
$M_{B}$ & 142 $\pm$ 11 $M_{Jup}$\tablenotemark{a}  \\
$M_{B}$ & 0.14 $\pm$ 0.01 $M_{\sun}$\tablenotemark{a}  \\
\enddata
\tablenotetext{a}{Assuming $M_{A} = 1.02 \pm 0.03 M_{\sun}$ \citep{takeda07}.}
\label{tab:table7}
\end{deluxetable}

\clearpage

\clearpage

\clearpage

\clearpage

\clearpage

\clearpage

\end{document}